\documentclass[%
 aps,
 prx,
 reprint,
 superscriptaddress,
 amsmath,amssymb,
floatfix,
]{revtex4-2}

\usepackage[english]{babel}
\usepackage{graphicx}
\graphicspath{{figures/}}

\usepackage{dcolumn}
\usepackage{bm}
\usepackage{braket}

\usepackage[
range-units = single,
retain-zero-uncertainty = true,
per-mode = symbol,
]{siunitx}
\DeclareSIUnit{\gauss}{G}

\usepackage{xspace}
\usepackage{isotope}
\newcommand{\calcium}{$\isotope[40]{Ca}^+$\xspace}

\usepackage{dsfont}

\usepackage{xcolor}
\usepackage[colorlinks=true, allcolors=blue]{hyperref}

\usepackage{verbatim}
\usepackage{upgreek}

\usepackage{xstring}
\newcommand*{\aref}[1]{%
	\IfBeginWith{#1}{eq:}{Eq.~\eqref{#1}}{}%
	\IfBeginWith{#1}{fig:}{Fig.~\ref{#1}}{}%
	\IfBeginWith{#1}{tab:}{Table~\ref{#1}}{}%
	\IfBeginWith{#1}{appendix:}{Appendix~\ref{#1}}{}%
	\IfBeginWith{#1}{sec:}{Section~\ref{#1}}{}%
}

\usepackage{subfiles}

\newcommand{\thetitle}{
    State-dependent control of the motional modes of trapped ions using an integrated optical lattice
}
\newcommand{\affiliationTiqi}{
\affiliation{Institute for Quantum Electronics, ETH Z\"{u}rich, 8093 Z\"{u}rich, Switzerland}
}

\newcommand{\theauthors}{

\author{Alfredo Ricci Vasquez}
\email{aricci@ethz.ch}
\thanks{These two authors contributed equally}
\affiliationTiqi

\author{Carmelo Mordini}
\thanks{These two authors contributed equally}
\affiliationTiqi

\author{Daniel Kienzler}
\affiliationTiqi

\author{Jonathan P. Home}
\email{jhome@ethz.ch}
\affiliationTiqi
\affiliation{Quantum Center, ETH Z\"{u}rich, 8093 Z\"{u}rich, Switzerland}

}

\begin{document}

\title{\thetitle}
\date{\today}
\theauthors

\begin{abstract}

In this work we study the interaction of trapped ions with a state-dependent, high-intensity optical lattice formed above an ion trap chip using integrated photonics. We use a single ion to map the optical potential landscape over many periods of the standing-wave field.  For a single ion sitting in the centre of the lattice we observe a state-dependent trap-frequency shift of $2\pi\times\SI{3.33(4)}{\kilo\hertz}$, corresponding to a bare optical potential of $2\pi\times\SI{76.8(5)}{\kilo\hertz}$ for the electronic ground state. We extend this to two ions, measuring state-dependent shifts of both axial modes. Additionally, using the internal-state dependence of the interaction, we perform a direct measurement of the energy distribution of the motion of a single ion using carrier spectroscopy. Improvements to the setup would allow to increase the state-dependent curvature by more than 50 times, providing a tool which can be utilised for motional state control, and multi-ion gates using optical potentials produced in a scalable fashion. 

\end{abstract}


\maketitle

Optical dipole potentials play a central role in the quantum control of atomic systems. For neutral atoms and particles, these potentials play a dual role both for trapping as well as for internal-state dependent control  \cite{gross_quantum_2021, kaufman_quantum_2021, schneider_optical_2010, enderlein_single_2012, lambrecht_long_2017, schmidt_optical_2018}. In particular, the use of atoms trapped in optical lattices has allowed exploration of a rich variety of physics, particularly in the domain of quantum simulation \cite{bloch_quantum_2012, kaufman_quantum_2021}. For trapped ions, where strong electric trapping is available, internal state dependent optical dipole forces are the main tool used for multi-qubit gates \cite{ blatt_entangled_2008, leibfried_experimental_2003, clark_high-fidelity_2021, sawyer_wavelength-insensitive_2021}. These have commonly been implemented using travelling wave fields, in which the force Hamiltonian is picked out as one resonance among many, although using positioning of the ion in standing waves may offer access to higher speeds \cite{mehta_fast_2019, saner_breaking_2023, ricci_standing-wave_2023}. Higher order derivatives of the optical potential can produce state-dependent modifications to the ions' motional frequencies, producing a Hamiltonian $\propto \hat\sigma_z \hat a^\dagger \hat a$ which has for a long time been at the core of oscillator state control for cavity-QED systems \cite{brune-1990, guerlin_progressive_2007, lee-2019, schuster_resolving_2007, arrangoiz-arriola_resolving_2019}.
Such Hamiltonians have have recently been proposed as a tool for realising entangling gates for ions using global electric fields \cite{mazzanti_trapped_2021, mazzanti_trapped_2023} and performing state-dependent squeezing \cite{shapira_robust_2023, drechsler_state-dependent_2020}. They have also been used for purifying large ion crystals \cite{weckesser_trapping_2021} as well as for laser cooling \cite{taieb-cooling-1994, cooper_alkaline-earth_2018}. 
The possibility of precisely controlling the position of ion crystals in high-intensity optical lattices has also been proposed for measuring parity violation using ions \cite{koerber_radio_2003, fortson_possibility_1993}, and used for the study of friction models \cite{Bylinskii2015, Pruttivarasin2011, Garcia-Mata2007}.
Engineering motional modes using optical potentials lies at the centre of several proposals to perform quantum control with large crystals of ions \cite{schwerdt_scalable_2023, arias_espinoza_engineering_2021, olsacher_scalable_2020, teoh_manipulating_2021}. However while trapping of ions in optical lattice potentials has been realized \cite{enderlein_single_2012, hoenig_trapping_2024}, the state-dependent curvature of a standing-wave optical potential has not yet been observed experimentally.

In this Letter we demonstrate state-dependent motional frequencies of a single and two-ion crystals, by positioning them with sub-wavelength accuracy in a high-intensity optical lattice \cite{Linnet2012, begley_optimized_2016, laupretre_controlling_2019} generated using integrated photonics \cite{Mehta2016, Niffenegger2020, Ivory2020, Mehta2020, ricci_standing-wave_2023, kwon_multi-site_2023}.
By locating the ion in different positions of the lattice we observe modifications of the motional frequencies of the ion(s).
We investigate the state-dependent nature of this potential by performing measurements for two different electronic states, and find good agreement with theoretical values. We use this state-dependent potential to directly observe the thermal distribution of the motion of the ion through spectroscopy of the qubit \cite{meir-2017}. The use of integrated photonics guarantees excellent passive phase stability and power distribution between the beams forming the lattice with minimal technical overhead \cite{Mundt2002, Schmiegelow2016, saner_breaking_2023, wipfli_integration_2023}, and could also in future allow for incorporation of optical lattice potentials in large scale architectures \cite{kielpinski_architecture_2002, mordini_multi-zone_2024, kwon_multi-site_2023} for both atoms and ions. 

The lattice is created using light at $\lambda\approx \SI{733}{\nano\meter}$ from the counter-propagating components of two beams emitted at an angle of $\theta\approx36^{\text{o}}$ from two waveguide outcouplers in our surface-electrode ion trap (see \aref{fig:first} (a)). The two beams are sourced from a single fiber input, and split on chip using an integrated waveguide splitter near the trapping location, providing a phase stable configuration which is stably referenced to the ion position \cite{ricci_standing-wave_2023, Mehta2020}.
The \SI{733}{\nano\meter} light of the optical lattice primarily couples off-resonantly  to the dipole-allowed transitions at $\SI{393}{nm}$, $\SI{397}{nm}$ and $\SI{854}{nm}$ in the \calcium ion. This coupling results in an ac Stark-shift of the chosen qubit states  $\ket{\downarrow} = \ket{4S_{1/2}, m_j = -1/2}$ and $\ket{\uparrow} = \ket{3D_{5/2}, m_j = -1/2}$, connected by a quadrupole transition at $\SI{729}{\nano\meter}$ \cite{ricci_standing-wave_2023, Haffner2003}.
\aref{fig:first} (b) shows the relevant energy levels and transitions of the \calcium ion.
The dashed lines represent the shifted energy levels when applying light at 733 nm with intensity $I(x)$ to the ion. 
Explicitly, this shift is given by $\Delta E_i = \hbar\alpha_iI(x)/2c\epsilon_0$ with $i\in\{\uparrow, \downarrow\}$, and $\alpha_i$  the polarisability of the $\ket{i}$ state for light at $\lambda$ \cite{delone_ac_1999}.
The ac Stark-shift is described by the Hamiltonian $\hat{H} = \hbar(\gamma\hat{\sigma}_z +\beta\mathds{1})I(x)$, where $\gamma = (\alpha_\uparrow- \alpha_\downarrow)/4c\epsilon_0=2\pi\times\SI{4.37e-4}{\hertz/(\W/\meter^2)}$ , $\beta = (\alpha_\uparrow+ \alpha_\downarrow)/4c\epsilon_0 = -2\pi\times\SI{0.65e-4}{\hertz/(\W/\meter^2)}$ and $\hat\sigma_z = \ket{\downarrow}\bra{\downarrow}-\ket{\uparrow}\bra{\uparrow}$ the Pauli z-operator.

Additional to the integrated beams used for the optical lattice, we use a low-intensity free-space beam at $\SI{729}{nm}$ directed along the trap axis to near-resonantly drive the $\ket{\downarrow}\leftrightarrow\ket{\uparrow}$ transition. We also use two free-space \SI{397}{\nano\meter} beams for Doppler and Electromagnetically-Induced Transparency (EIT) cooling \cite{lechner_electromagnetically-induced-transparency_2016}, as well as for preparing the $\ket{\downarrow}$ state by optical pumping. Finally, we use two integrated beams at \SI{866}{\nano\meter} and \SI{854}{\nano\meter} for repumping \cite{Haffner2008}. Further information about the experimental apparatus and the pulse sequences can be found in the supplementary material (SM) \footnote{Supplementary material can be found in ...} and Ref. \cite{ricci_standing-wave_2023}. 

\begin{figure}[t!]
    \centering    \includegraphics[width=\columnwidth]{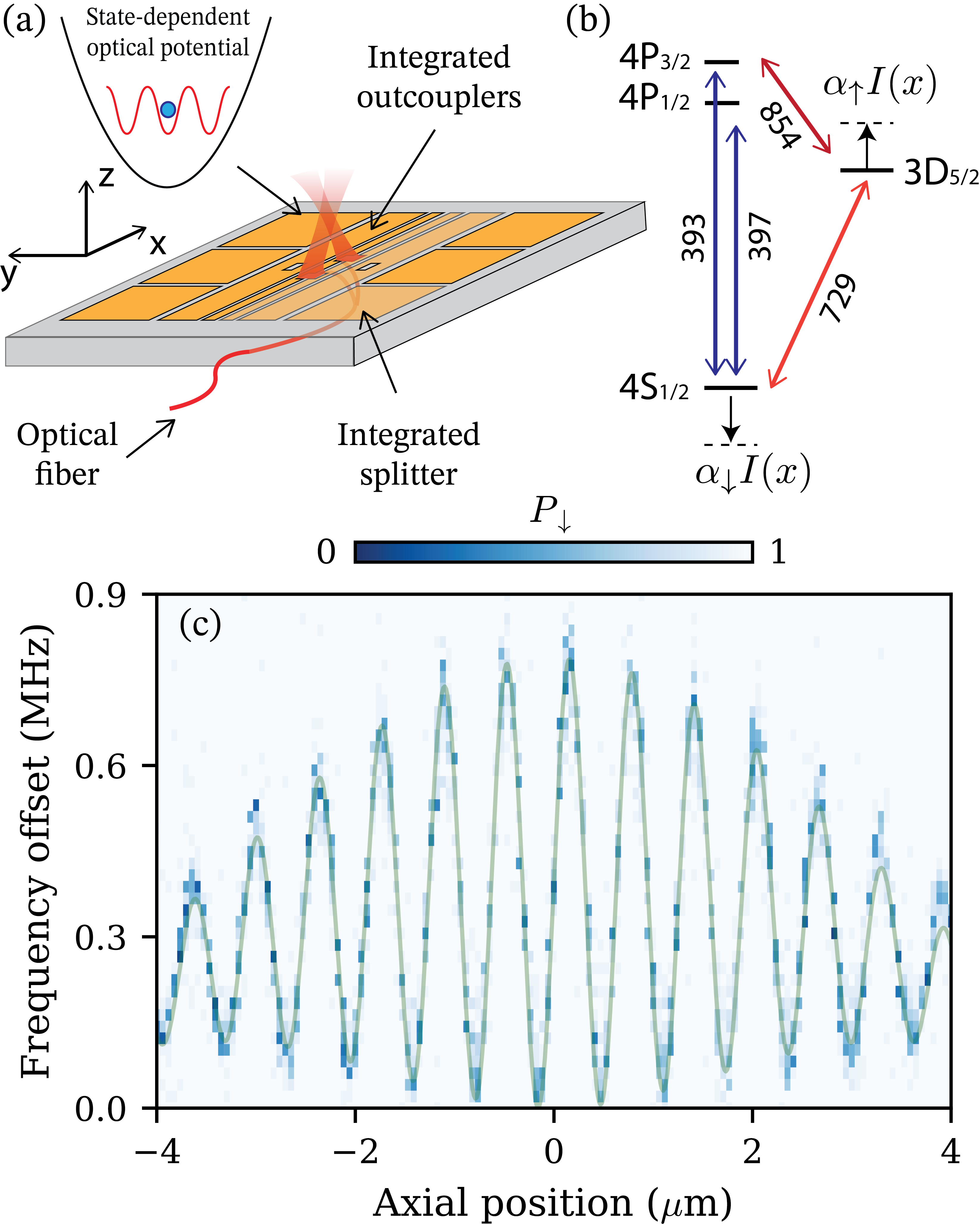}
    \caption{\textbf{Experimental layout and optical potential landscape.} (a) Sketch of the trap. Integrated waveguides are used to route the light from optical fibers to grating outcouplers. The lattice is generated from two counter-propagating beams emitted from the trap. (b) Relevant energy levels of the \calcium ion as well as the wavelengths of its transitions. The optical lattice couples off-resonantly to the dipole transitions at \SI{393}{\nano\meter}, \SI{397}{\nano\meter}, \SI{854}{\nano\meter} shifting the energies of the $\ket{\downarrow}$ and $\ket{\uparrow}$ states. (c) Optical potential landscape of the lattice measured using the $\SI{729}{\nano\meter}$ beam. The green line serves as a guide for the eye.}
    \label{fig:first}
\end{figure}

With a single single ion we map out the intensity landscape of the optical lattice \cite{guthohrlein_single_2001}, using $\approx\SI{200}{\milli\watt}$ of power coupled into the fibers routing the light to the chip.
Assuming total losses from the fibers, waveguides, and outcouplers, which we estimated using independent measurements to be $\approx \SI{7}{\decibel}$, we expect the lattice to have a total power of $\approx \SI{40}{\milli\watt}$.
We perform qubit spectroscopy using a finite-duration probe pulse of the $\SI{729}{nm}$ beam while the optical lattice is on.
We repeat this measurement over a range of $\pm\SI{4}{\micro\meter}$ along the trap axis.
\aref{fig:first} (c) shows the subsequent population left in the $\ket{\downarrow}$ state as a function of the axial position of the ion and the detuning of the $\SI{729}{nm}$ laser from the bare qubit frequency $\omega_0$.
Assuming that the frequency shift of the qubit originates solely from the lattice-induced differential ac Stark-shift $\Delta_z$, these measurements allow us to retrieve the intensity pattern $I(x)=\Delta_z(x)/2\gamma $.

The curvature of the ac Stark-shift Hamiltonian induces position- and state-dependent changes in the motional frequency of the ion(s).
Ignoring the curvature arising from the beam profile, we can approximate the state-dependent trap frequency of the optical potential along the trap axis as

\begin{equation}
    \omega_i^2(x_0) = -\frac{2\hbar\Delta'_zk^2_xr_i}{m}\cos(2k_x(x_0-x')),
    \label{eq:omega_i}
\end{equation}
where $x_0$ is the centre of motion of the ion, $r_i=\alpha_i/(\alpha_\uparrow-\alpha_\downarrow)$ the ratio of the polarisability of the $\ket{i}$ state to the differential polarisability, $x'$ is a position offset, $k_x = 2\pi\sin{\theta}/\lambda$ the x-component of the wavevector, and $\Delta'_z$ is the measured \textit{local} minimum-to-maximum amplitude of $\Delta_z$ for the fringe where the ion is located, and $m$ the \calcium mass.
When this potential gets added to the electric potential of the ion with frequency $\omega_x$, the state-dependent motional frequency of the ion becomes $\omega^i_x = (\omega_x^2+\omega^2_i)^{1/2}$. Assuming $\omega_x\gg\omega_i$, this becomes

\begin{equation}
     \omega^i_x(x_0)\approx \omega_x - 2\Delta'_zr_i\eta^2\cos(2k_x(x_0-x')),
    \label{eq:trap-frequencies}
\end{equation}
with $\eta = k_x\sqrt{\hbar/2m\omega_x}$ the Lamb-Dicke parameter for the optical lattice. We can maximise the second term in the expression by locating the ion in the node or anti-node of the standing wave (i.e. $2k_x(x_0-x') = n
\pi$ with $n \in \mathcal{Z}$). Restricting the analysis to this case and neglecting the common frequency shift proportional to $\beta$, we find the effective Hamiltonian of the interaction of the ion with the lattice to be,

\begin{equation}
    \hat{H}= \hbar\omega_t\hat\sigma_z(\hat a^\dagger \hat a+1/2),
    \label{eq:hamiltonian}
\end{equation}
where $\omega_t = 2\eta^2\Delta'_z = \omega^\uparrow_x-\omega^\downarrow_x$ is the differential optical confinement strength between the qubit states. Here, we have ignored the state-dependent squeezing terms $(\hat{a}^\dagger)^2+(\hat{a})^2$ resonant with twice the trap frequency, since the optical lattice is kept at a constant power. It would be possible, however, to modulate the optical lattice in order to achieve state-dependent squeezing \cite{drechsler_state-dependent_2020, shapira_robust_2023}.

\begin{figure}[t!]
    \centering
    \includegraphics[width=\columnwidth]{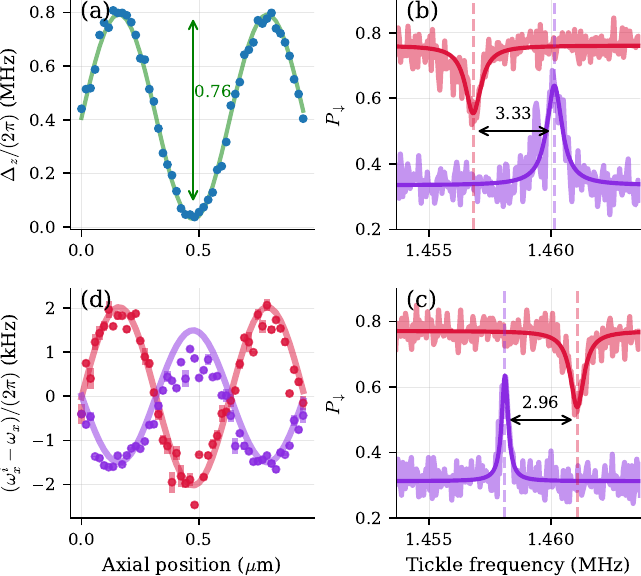}
    \caption{\textbf{State-dependent motional frequency for a single ion.} (a) Ac Stark-shift as a function of the axial position, measured doing qubit spectroscopy. The green line is a sinusoidal fit to the data.
    (b,c) Ground-state tickling for for an ion in the node (b) and anti-node (c) of the optical lattice. Red and purple lines are for an ion initialised in the $\ket{\downarrow}$ and $\ket{\uparrow}$ states.
    (d) Change in motional frequency for both qubit states as a function of position. The red and purple lines are calculated from theory following Eq. \ref{eq:trap-frequencies} and the fit from (a).}
    \label{fig:sinlge-ion}
\end{figure} 

 We investigate the state-dependent potential by measuring the axial motional frequency of the ion as a function of its axial position for positions covering one period in the center-most fringe of the optical lattice.
 We measure the state-dependent trap frequency $\omega^i_x$ by performing resonant driving of an ion prepared near the ground state of motion.
 We apply a low amplitude RF pulse to one of the DC electrodes, which when resonant with the motional frequency creates a coherent state.
 The excitation pulse is followed by a red sideband probe pulse using the \SI{729}{\nano\meter} beam. Here, the red sideband pulse will only induce a spin flip $\ket{\downarrow, n} \rightarrow \ket{\uparrow, n-1}$ when the ion is not in the motional ground state.
 For an ion prepared in the $\ket{\uparrow}$ state, we apply a similar sequence using instead a blue sideband, driving the $\ket{\uparrow, n} \rightarrow \ket{\downarrow, n-1}$ transition.
 This method, which we refer to as ground-state tickling, allows us to measure the trap frequency for low amplitude coherent states. This is essential for keeping the ion well-localised in the anharmonic optical lattice \cite{home_normal_2011}.
 The contrast of the resulting signal is limited by the finite temperature of the ion after EIT cooling. 

 \aref{fig:sinlge-ion} shows the state-dependent motional frequency of a single ion. 
  \aref{fig:sinlge-ion} (a) shows $\Delta_z$ as a function of the axial position of the ion, in the fringe of interest, extracted from a spectroscopy measurement similar to the one in \ref{fig:first} (c). 
 The green line corresponds to a fit of the ac Stark-shift using the form $\Delta_z(x_0)=\Delta'_z\cos^2(k_x(x_0-x'))+\Delta_0$, from which we extract $\Delta'_z = 2\pi\times\SI{0.76(1)}{\mega\hertz}$, $x'$, $k_x$ and the offset $\Delta_0$. This measurement allows us to estimate the maximum intensity of the lattice $I_0=\Delta'_z/2\gamma
 = \SI{0.88(1)}{\giga\watt\per\meter\squared}$.
 \aref{fig:sinlge-ion} (b) and (c) show the population in $\ket{\downarrow}$ after performing ground state tickling for an ion sitting in the node and anti-node, respectively, and for both qubit states. For these experiments we set the bare trapping frequency to $
\omega_x = 2\pi\times 
\SI{1.4578}{\mega\hertz}$, resulting in $\eta = 0.048$. A dip or peak in the population occurs when the tickling frequency matches the frequency of the confining potential, allowing us to measure $\omega^i_x$.
Using Lorentzian fits we extract a differential state-dependent optical confinement $\omega_t = 2\pi\times\SI{3.33(4)}{\kilo\hertz}$ for the ion located in the node and of $\omega_t = -2\pi\times\SI{2.96(2)}{\kilo\hertz}$ when the ion is located in the anti-node.
 We find these values to be lower than the maximum expected value of $|\omega_t| = |2\eta^2\Delta'_z| = 2\pi\times\SI{3.52(5)}{\kilo\hertz}$, and we attribute this discrepancy,
 to changes in $\omega_x$ arising from stray potentials caused by changes in the duty cycle of the high-power beam \cite{malinowski_generation_2022}.
\aref{fig:sinlge-ion} (d) shows the motional frequency offset $\omega^i_x-\omega_x$ as a function of axial position, for an ion prepared in the $\ket{\downarrow}$ and $\ket{\uparrow}$ states.
The data points are extracted from Lorentzian fits similar to those in Figures \aref{fig:sinlge-ion} (b, c).
The lines are calculated from theory following \aref{eq:trap-frequencies} and using $\Delta'_z$, $x'$ and $k_x$ from the ac Stark-shift fit.
We found the maximum absolute change in the trapping frequency ${\left|2r_i\Delta'_z\eta^2\right|}$ to be $2\pi\times\SI{2.02(3)}{\kilo\hertz}$ and $2\pi\times\SI{1.50(2)}{\kilo\hertz}$, corresponding to maximum absolute bare optical trap frequencies of $2\pi\times\SI{76.8(5)}{\kilo\hertz}$ and $2\pi\times\SI{66.1(4)}{\kilo\hertz}$ for the $\ket{\downarrow}$ and $\ket{\uparrow}$ states.
  
The Hamiltonian of \aref{eq:hamiltonian} predicts a phonon-dependent ac Stark-shift of the qubit frequency, as well as a qubit energy-state dependent change in the trap frequency. When the ion is located in the node, this shift happens because higher Fock states sample regions of higher intensities of the lattice. For an ion located in the anti-node of the lattice the opposite happens. The dressed frequency of the qubit is thus dependent on the occupied Fock state $\ket{n}$ as $\omega_0\rightarrow\omega_0+(1/2+n)\omega_t$ (see SM for details). This effect allows us to measure the phonon distribution of the ion using qubit spectroscopy. We demonstrate this by preparing the ion in a thermal state and performing spectroscopy of the qubit using the 
\SI{729}{\nano\meter} beam while the ion sits in the node of the optical lattice.
We apply the spectroscopy pulse with a duration $t_{\pi}=\SI{352}{\micro\second}$, which corresponds to a $\pi$-pulse on the qubit when the lattice is off. For this experiment we independently measure $\Delta_z = 2\pi\times\SI{0.281(3)} {\mega\hertz}$ and a trap frequency of $\omega_x =2\pi\times\SI{1.8452}{\mega\hertz}$ leading to $\omega_t=2\pi\times\SI{0.98(1)}{kHz}$. 
The results are shown in \aref{fig:distributions} (a), where the lines are fits to 

\begin{equation}
    \label{eq:thermal_carrier}
    P_\downarrow = \sum_{n=0}^{\infty} \mathcal{P}(n) \frac{\Omega}{\Omega_{\text{eff}}} \cos^2(\Omega_{\text{eff}}t_\pi / 2).
\end{equation}
Here, $\Omega = \pi/t_\pi$ is the Rabi frequency of the \SI{729}{\nano\meter} beam, $\Omega^2_{\text{eff}} = \Omega^2 + (\delta - ((1/2)+n)\omega_t-\Delta_0)^2$ the effective Rabi frequency for Fock state $\ket{n}$, and $\delta$ the detuning between the \SI{729}{\nano\meter} beam and the bare qubit frequency. $\mathcal{P}(n)$ is the distribution of the Fock states, where we consider both a thermal and a displaced-thermal distribution \cite{ruster_experimental_2014} (see SM for more details).
From fitting a thermal state we infer a mean phonon occupation $\bar n = 6.4(3)$, and an extinction ratio of the intensity $\Delta_0/\Delta'_z$ below $1.6\%$.
For a fit using a displaced thermal state we find the thermal occupation $\bar {n} = 1.4(2)$, a displacement amplitude $|\alpha| = 2.21(8)$ and an extinction ratio of $1.2 \%$. 
While the theory reproduces the general trend of the measurement over the entire frequency range, both fits show discrepancies with the data which we attribute to the ion heating up during the pulse. 
Our method can in principle resolve single Fock states, however this would require longer probe times or larger values of $\omega_t$.
The former is unfeasible due to the large axial heating rates of  $2-3$ quanta/ms observed in our trap \cite{Mehta2020}.
The latter would require increasing the light intensity at the ion or using wavelengths with larger differential polarisability of the qubit states. We discuss the prospects of doing this in the SM.

This method provides a useful direct diagnostic of the ion's motional energy distribution. We test this method by preparing the motion of the ion at different temperatures by changing the detuning of the laser beams used for sub-Doppler EIT cooling, and repeat similar experiments as in \aref{fig:distributions} (a) \cite{lechner_electromagnetically-induced-transparency_2016}. The results are shown in \aref{fig:distributions} (b).
The data show dependency of the spread of the spectroscopy signal on EIT detuning, which result from differences in the temperature of the ion. 
We verify this dependency by performing a temperature characterisation using a blue sideband (BSB) probe pulse after EIT cooling. 
\aref{fig:distributions} (c) shows the first cycle of thermal BSB Rabi oscillations as a function of the EIT detuning. Slower BSB Rabi oscillations correspond to a colder ion.
The regions where the BSB is the slowest coincides with the direct carrier spectroscopy of the thermal state having a narrow distribution close to the bare qubit frequency. 
For frequencies where the BSB is the fastest we observe that the spectroscopy signal has the highest spread.
The two methods show good agreement, with the first method being advantageous for reconstructing the Fock state distribution since it directly reveals the phonon distribution in a single frequency sweep. In contrast, for the blue-sideband a full reconstruction would require taking time-scans over many cycles of Rabi oscillation and extracting the weights of multiple frequency components \cite{meekhof_generation_1996}.

\begin{figure}[ht!]
    \centering
    \includegraphics[width=\columnwidth]{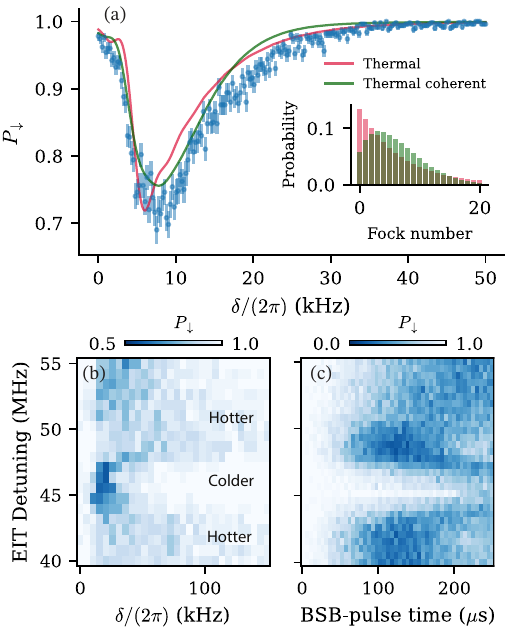}
    \caption{\textbf{Thermometry of a single ion using the state-dependent optical potential.} (a) Qubit spectroscopy using the 729 beam for an ion located in the node of the lattice. 
    The red and green lines are fits to Eq. \ref{eq:thermal_carrier} using thermal and displaced thermal  distributions.
    The inset shows $\mathcal{P}(n)$ for both fits. (b) Qubit spectroscopy similar to (a) as a function of the detuning of the EIT cooling beams.
    (c) Rabi oscillations of the BSB, as a function of the detuning of the EIT cooling beams.}
    \label{fig:distributions}
\end{figure}

In a second set of experiments, we investigated state-dependent optical potentials for a two-ion crystal, which has been suggested as an ingredient to  perform new types of multi-qubit trapped-ion gates \cite{mazzanti_trapped_2021, mazzanti_trapped_2023}.
We operate in a configuration were both ions are simultaneously located either in a node or anti-node of the lattice, separated by $j=3$ periods.
In this configuration the theoretical trap frequency of the center-of-mass (COM) mode is $2\pi\times\SI{1.83}{\mega\hertz}$ and the ion separation $d=\SI{3.74}{\micro\meter}$.
Other values of $j$ are in principle possible, but result in reduced Lamb-Dicke parameters or reduced lattice intensities at the ion positions. 
We can measure the ac Stark-shift on both ions as a function of the axial position of the crystal.
We do this by measuring the $\ket{\downarrow\downarrow}$ population as a function of the equilibrium position of the crystal and the frequency offset of the \SI{729}{\nano\meter} beam, similarly to the single ion experiment of \aref{fig:first} (c).
We repeat this measurement and fine-tune the trap frequency until the absorption signals of both ions overlap  (see the SM for further details).
The resulting signal is shown in \aref{fig:two-ions} (a).
From the data we extract $\Delta'_z=2\pi\times\SI{0.57(1)}{\mega\hertz}$, $x'$ and $k_x$ using a sinusoidal fit.
We find the COM trap frequency for which the ions have even intensities to be $\omega^{\text{com}}_x = 2\pi\times\SI{1.8485}{\mega\hertz}$ with Lamb-Dicke parameters of $\{\eta_{\text{com}}, \eta_{\text{str}}\} = \{4.17, 3.17\}\times10^{-2}$ for the COM and out-of-phase (STR) mode. 
We then proceed to measure the state-dependent trapping frequencies $\omega^{i}_x$ with $i \in\{\downarrow\downarrow, \uparrow\uparrow\}$ for the two motional modes, as a function of the equilibrium position of the ions in the lattice.
\aref{fig:two-ions} (b), (c) show the measurement of the state-dependent motional frequency $\omega^{i}_x-\omega_x$ for the COM and STR mode.
The data points are extracted from absorption signals similar to \aref{fig:sinlge-ion} (b, c).
The lines are generated from theory using \aref{eq:omega_i} and the fitted values for the ac Stark-shift.
We find the maximum state-dependent change of the trap frequencies $|2\eta_{\text{com}}\Delta'_zr_i|$ to be $2\pi\times\SI{1.15(1)}{\kilo\hertz}$ and $2\pi\times\SI{0.85(1)}{\kilo\hertz}$ for the $\ket{\uparrow\uparrow}$ and $\ket{\downarrow\downarrow}$, respectively. For the STR mode we find $|2\eta_{\text{str}}\Delta'_zr_i|$ to be $2\pi\times\SI{0.66(1)}{\kilo\hertz}$ and $2\pi\times\SI{0.49(1)}{\kilo\hertz}$ in good agreement with the theoretical values. 

\begin{figure}
    \centering
    \includegraphics[width=\columnwidth]{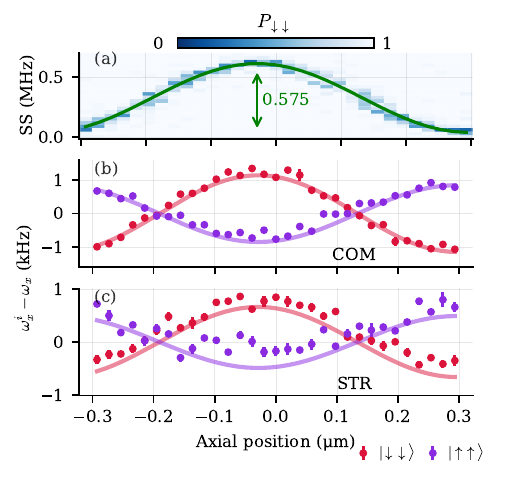}
    \caption{\textbf{Motional frequency shifts of a two ion crystal for different axial positions. }(a) Population in the $\ket{\downarrow\downarrow}$ state as a function of the 729 laser detuning and the equilibrium position of the crystal. The green line is a sinusoidal fit to the data. (b,c) State-dependent motional frequency change as a function of axial position for the COM mode and STR mode, respectively. For the red and purple data points the ion is initialised in the $\ket{\downarrow\downarrow}$ and $\ket{\uparrow\uparrow}$ states. The solid lines are calculated from theory, using the fit from (a).}
    \label{fig:two-ions}
\end{figure}

Our results highlight the potential of creating high-intensity light fields sourced from integrated photonics, while keeping the ion well localised.
By choosing a more suitable wavelength of the lattice for which the differential polarisability of the states is higher, it should be possible to increase the curvature of the potential. For calcium this could be done by operating at wavelengths closer to \SI{854}{\nano\meter} or \SI{397}{\nano\meter}, allowing to increase the differential optical trapping strength by more than an order of magnitude, and thus reducing the need to operating at high powers (details regarding the attainable shifts and photon scattering rates are given in the SM). This would allow resolving single Fock states for motional state readout, opening up quantum control using the state-dependent shift, while the extension to two ions would also enable novel trapped-ion multi-qubit gates \cite{mazzanti_trapped_2021}. Compared to other methods which were able to obtain energy level dependent internal state shifts in trapped ions, our technique benefits from a relatively simple implementation and lack of off-resonant terms \cite{ding_cross-kerr_2017, roos_nonlinear_2008, schmidt-kaler_quantized_2004, mallweger_single-shot_2023}. Wavelengths near \SI{397}{\nano\meter} would allow optical trapping frequencies of the electronic ground state of hundreds of kHz, also opening the possibility for on-chip optical trapping. This could provide the possibility to combine ion and neutral atoms trapping on chip, opening up a new platform for the study of atom-ion interactions \cite{lambrecht_long_2017, pinkas_trap-assisted_2023}. The possibility to use integrated optics to combine tight focussing \cite{beck_grating_2024} with phase stable optical lattices has potential to go beyond standard free-space paradigms, opening new avenues for both atoms and ions.

We acknowledge support from the Swiss National Science Foundation (SNF) under Grant No. 200020\_207334 and the Intelligence Advanced Research Projects Activity (IARPA), via the US Army Research Office (ARO) under Cooperative Agreement Number W911NF-23-2-0216 as part of the Entangled Logical Qubits (ELQ) program. We acknowledge  LioniX International for fabrication of the trap devices. 

We thank Karan Mehta, Maciej Malinowski and Chi Zhang for their contributions to the ion trap setup, and Silvan Koch for setting up the Ti:Sa laser and providing us with technical support for its use. We thank Wojtek Adamczyk for his support in the use of the Atomphys library used for the calculation of the polarisabilities. We thank Tanja Berhle, Henrik Hirzler and Felix Knollmann for helpful comments of the manuscript. ARV also thanks Wojtek Adamczyk and Roland Matt for stimulating discussions.

ARV, CM and JH devised the experiments, which were carried out by ARV and CM. The paper was written by ARV and JH with input from all authors. The work was supervised by DK and JH.

\bibliography{bibliography}

 \renewcommand{\bibliography}[1]{}  

\setcounter{equation}{0}
\setcounter{figure}{0}
\setcounter{table}{0}
\setcounter{page}{1}
\makeatletter
\renewcommand{\theequation}{S\arabic{equation}}
\renewcommand{\thefigure}{S\arabic{figure}}
\renewcommand{\bibnumfmt}[1]{[S#1]}
\makeatother

\title{Supplemental materials for:\\\thetitle}
\theauthors

\maketitle

\section{Extended theory}

\subsection{Interaction of the ion with the lattice}
\label{sec:extended-theory-1}

We begin by describing the interaction of a single ion with the optical lattice \footnote{For simplicity we will use $
\hbar=1$ during the supplementary material}. We model the two beams emitted from the trap as two plane waves counter-propagating in the axial direction $x$, and co-propagating in the out-of-plane direction $z$. 
The light emitted from the two couplers can be approximated as,

\begin{equation}
\begin{aligned}
    \mathbf{E}(x,z) = E_0e^{k_xx+k_zz}\mathbf{e}_y+E_0e^{k_zz-k_xx}\mathbf{e}_y\\ = E_0e^{ik_zz}\cos(k_xx)\mathbf{e}_y,
\end{aligned}
\end{equation}
with $k_x$ as defined in the main text, and $k_z$ the projection of the wavevector in the $z$ direction. Here we assume that both beams have equal amplitudes $E_0$. This field results in an intensity landscape given by

\begin{equation}
    I(x) = \frac{\epsilon_0c|\mathbf{E}(x,z)|^2}{2} = I_0\cos^2(k_xx),
    \label{eq:intensity}
\end{equation}
with $I_0 = c\epsilon_0|E_0|^2/2$ the maximum intensity. Since this field is far-detuned from any relevant transition, we describe the coupling between the lattice and the ion in the ion interaction frame as
\begin{equation}
    \hat H_{\text{int}}=
    \begin{pmatrix}
        \Delta E_\uparrow & 0\\
        0 & \Delta E_\downarrow
    \end{pmatrix},
\end{equation}
where

\begin{equation}
    \Delta E_i =  \frac{\alpha_i(\lambda, \phi)|E|^2}{4} = \frac{\alpha_i(\lambda, \phi)I(x)}{2c\epsilon_0},
\end{equation}
is the ac Stark shift of the $\ket{i}$ state with polarisability $\alpha_i(\lambda, \phi)$, depending both on the lattice wavelength and the angle between the magnetic field and the polarisation $\phi$ \cite{delone_ac_1999}.
 We find the polarisabilities of the qubit states to be $\alpha_\downarrow/(2c\epsilon_0) = -2\pi\times\SI{5.01e-4}{\hertz/(\W/\meter^2)}$ and $\alpha_\uparrow/(2c\epsilon_0) = 2\pi\times\SI{3.72e-4}{\hertz/(\W/\meter^2)}$. Direct calculation of these elements can be done following, for example, Ref. \cite{sawyer_wavelength-insensitive_2021}. The Hamiltonian can be re-expressed as

\begin{equation}
\hat H_{\text{int}}=\frac{I(x)}{2c\epsilon_0}
    \begin{pmatrix}
        \alpha_\uparrow & 0\\
        0 & \alpha_\downarrow
    \end{pmatrix},
\end{equation}
or alternatively,

\begin{equation}
    \begin{aligned}
    \hat H_{\text{int}} = \frac{I(x)}{4c\epsilon_0}\left[(\alpha_\uparrow-\alpha_\downarrow)\hat\sigma_z+(\alpha_\uparrow+\alpha_\downarrow)\mathds{1}\right]\\ = (\gamma\hat{\sigma}_z +\beta\mathds{1})I(x),
    \end{aligned}
\end{equation}
with $\gamma$ and $\beta$ as defined in the main text.
We consider the motion of the ion along the trap axis as a harmonic oscillator described by the Hamiltonian

\begin{equation}  
\hat{H}_m = \omega_x(\hat{a}^\dagger\hat a + 1/2),
\end{equation}
with $\omega_x$ the axial trap frequency, and $\hat a$ and $\hat a^{\dagger}$, the creation and annihilation operators of the harmonic oscillator. The position of the ion in the field can be expressed as $x = x_0 + \hat{x}$, where $\hat{x} =a_0(\hat{a}^\dagger+a)$ is the position operator of the ion in the harmonic potential centred around $x_0$, with $a_0 = \sqrt{\hbar/2m\omega_x}$ the size of the motional ground state of the ion. Expanding $\hat H_{\text {int}}$ to second order around $x_0$ gives

\begin{equation}
    \hat H_{\text {int}} \approx (\gamma\hat{\sigma}_z +\beta\mathds{1})\left[ I(x_0) + \partial_xI(x)|_{x_0}\hat{x}+(1/2)\partial^2_xI(x)|_{x_0}\hat{x}^2\right].
\end{equation}

Here, the first term $\hat H_{\text{ss}} = (\gamma\hat{\sigma}_z +\beta\mathds{1})I(x_0)$ is simply the motion-independent ac Stark-shift, dependent on the equilibrium position of the ion. 
For the second and third term we use the transformation into the interaction picture with respect to the harmonic oscillator ($\hat a \rightarrow \hat ae^{-i\omega_x t}$; $\hat a^\dagger \rightarrow \hat a^\dagger e^{i\omega_x t}$ ). The second term corresponds to a state-dependent force in the $\hat\sigma_z$ basis, 

\begin{equation}
\begin{aligned} 
    \hat H_{\text{force}} =(\gamma\hat{\sigma}_z +\beta\mathds{1})\partial_xI(x)|_{x_0}\hat{x},
\end{aligned} 
\end{equation}
which can be expressed as

\begin{equation}
\begin{aligned} 
    \hat H_{\text{force}} = -k_xI_0(\gamma\hat{\sigma}_z +\beta\mathds{1})\sin({2k_xx_0})\hat{x}\\
    = -\eta I_0(\gamma\hat{\sigma}_z+\beta\mathds{1})\sin({2k_xx_0})(\hat a e^{-i\omega_xt} + \hat a^\dagger e^{i\omega_xt}).  
\end{aligned} 
\end{equation}

 Here we introduce the Lamb-Dicke parameter for the lattice $\eta = k_xa_0$. This Hamiltonian is maximised for an ion sitting in the slope of the optical lattice. This term can be made resonant by modulating the intensity of the optical lattice at the the trap frequency. 
The next term is a state-dependent curvature, given by 

\begin{equation}
    \begin{aligned}
    \hat H_{\text{curv}} = (1/2)(\gamma\hat{\sigma}_z +\beta\mathds{1})\partial^2_xI(x)|_{x_0}\hat{x}^2\\
    =-k_x^2I_0(\gamma\hat{\sigma}_z +\beta\mathds{1})\cos(2k_xx_0)\hat{x}^2.\\
    \end{aligned}
    \label{eq:hamiltonian-curv}
\end{equation}

We use the commutator relation $[\hat a, \hat a^\dagger] = 1$ to expand the term $\hat x^2$, leading to,

\begin{equation}
    \begin{aligned}
    \hat{x}^2 = a^2_0(\hat a e^{-i\omega_xt}+ \hat a^\dagger e^{i\omega_xt})^2 \\
    = a^2_0(\hat a^2e^{-i2\omega_xt} + (\hat a^\dagger)^2e^{2i\omega_xt}+ 2\hat a^\dagger \hat a  +1).
    \end{aligned}
\end{equation}

As the intensity of the lattice is constant in time, we ignore the off-resonant terms $a^2e^{-i2\omega_xt} + (\hat a^\dagger)^2e^{2i\omega_xt}$ and we arrive to the state-dependent potential

\begin{equation}
    \hat H_{\text{curv}}
    =-2\eta^2I(x_0)\cos(2k_xx_0)(\gamma\hat{\sigma}_z +\beta\mathds{1})(\hat{a}^\dagger\hat{a}+1/2).\\
    \label{eq:state-dependent-h}
\end{equation}

We consider the case of an ion located either in the node or the anti-node, so that $\cos(2k_x x) = \pm1$, $\sin(2k_x x) = 0$, and ignore the common offset term $\beta$, arriving to the Hamiltonian given in the main text as  

\begin{equation}
    \begin{aligned}
    \hat H_{\text{curv}}
    =\pm2\eta^2\Delta'_z\hat{\sigma}_z(\hat{a}^\dagger\hat{a}+1/2)\\  =\pm\omega_t\hat{\sigma}_z(\hat{a}^\dagger\hat{a}+1/2),
    \end{aligned}
    \label{eq:hamiltonian-sd}
\end{equation}
where the sign is positive in the node and negative in the anti-node.

 We can extract the bare optical state-dependent trapping frequencies $\omega_i$ from  \aref{eq:hamiltonian-curv} using the relation,

\begin{equation}
    \frac{1}{2}m\omega^2_i\hat{x}^2 = \hat H_{\text{curv}},
\end{equation}
leading to,

\begin{equation}
\begin{aligned}
    \omega_i^2(x_0)
    =-\frac{I_0k^2_x\alpha_i}{mc\epsilon_0}\cos(2k_xx_0),
\end{aligned}
\end{equation}
or equivalently,

\begin{equation}
\begin{aligned}
    \omega_i^2(x_0)
    =-\frac{2\Delta'_zk^2_x}{m}\frac{\alpha_i}{\alpha_\uparrow-\alpha_\downarrow}\cos(2k_xx_0).
\end{aligned}
\end{equation}

We find the ratio of the polasability of the $\ket{i}$ state to the differential polarisabilites to be $r_i = \alpha_i/(\alpha_\uparrow-\alpha_\downarrow)$ = $\{-0.5744, 0.4256 \}$ for $i \in \{\downarrow, \uparrow\}$. Experimentally we can access the motional frequency of the ion under the electric and optical potential $\omega^i_x$,

\begin{equation}
\begin{aligned}
    \omega^i_x = (\omega_x^2+\omega^2_i(x_0))^{1/2}\\ \approx
    \omega_x +\frac{\omega^2_i(x_0)}{2\omega_x},\\
\end{aligned}
\end{equation}
leading to the equation reported in the main text,

\begin{equation}
\begin{aligned}
    \omega^i_x = \omega_x - \frac{2\Delta'_zk^2_x}{2m\omega_x}\frac{\alpha_i}{\alpha_\uparrow-\alpha_\downarrow}\cos(2k_xx_0)\\=
    \omega_x - 2\Delta'_z\eta^2\frac{\alpha_i}{\alpha_\uparrow-\alpha_\downarrow}\cos(2k_xx_0),
\end{aligned}
\end{equation}
in agreement with \aref{eq:state-dependent-h}.

\subsection{Validity of the model}

In the model described so far we assumed that the field at the ion location is created by two interfering plane waves with identical electric field amplitudes and phases.
This model can be extended to the real experimental conditions with minimal changes.
The main difference between the model and the experimental setup is the spatial profile of the beams forming the standing wave.
These beams have a Gaussian-like profile with a $1/e^2$ intensity radii at the focal point of $\sigma=\SI{6.5}{\micro\meter}$ along the $x$-direction and \SI{3.7}{\micro\meter} along the $y$-direction \cite{Mehta2020}. This will result in a reduced intensity outside the beams centre, which can be taken into account by fitting $I_0$ only for a single fringe.
The finite-size of the beams will also result in an additional gradient and curvature from the beam profile.
We ignore these contributions because $1/\sigma<<2k_x$.
Furthermore, the beam profile for each of the emitted beams is not completely symmetric along the trap axis, which will result in uneven electric field intensities from the beams outside their centre.
This results in a running-wave component in the $z$ direction, and therefore a non-vanishing intensity in the nodes of the optical lattice.
Finally, any path-length differences between the two arms result in a differential phase between the two beams.
In our data analysis we account for these differences by using a modified model where we assume that for each fringe of the lattice the intensity is,

\begin{equation}
    I(x) =  I'_0\cos^2(k_x(x_0-x')) +I_{\text{off}},
\end{equation}
with $I'_0$ the peak-to-peak intensity of the fringe, $I_{\text{off}}$ the intensity offset accounting for the running wave component and $x'$ the position offset. In terms of the experimental accessible ac Stark-shift $\Delta_z$ we get,

\begin{equation}
    \Delta_z(x) =  \Delta'_z\cos^2(k_x(x_0-x')) +\Delta_{0},
\end{equation}
with $\Delta_{0} = 2\gamma I_{\text{off}}$, the ac Stark-shift offset. The presence of a running-wave component results in a lower curvature of the optical potential which is accounted by a reduced $\Delta'_z$, and therefore the theory described in section \ref{sec:extended-theory-1} remains valid.

\subsection{Qubit spectroscopy of thermal states}

We use the state-dependent character of the optical lattice to measure the energy distribution of the motional state of the ion through the qubit frequency. When performing qubit spectroscopy of the qubit for an ion located in the node or anti-node of the optical lattice, the effective Hamiltonian for the interaction of the ion with both fields is

\begin{equation}
   \hat{H} = \hat{H}_{\text{sd}} + \hat{H}_{\text{axial}} + \hat{H}_\text{ss}, 
\end{equation}
with $\hat{H}_\text{ss} = \Delta_0\hat \sigma_z$ for an ion located in the node and $\hat{H}_\text{ss} = (\Delta_0+\Delta'_{z})\hat \sigma_z$ for an ion located in the anti-node. $\hat{H}_{\text{axial}}$ is the Hamiltonian for the interaction of the ion with the axial beam,

\begin{equation}
    \hat{H}_{\text{axial}} =  \frac{\Omega}{2} (\sigma_+ e^{-i \delta t} + \text{h.c.}), 
\end{equation}
where $\Omega$ is the carrier Rabi frequency, $\sigma_+ =\ket{\uparrow}\bra{\downarrow}$ the the atomic rising operator and $\delta=\omega_l-\omega_0$ the detuning between the \SI{729}{\nano\meter} beam frequency $\omega_l$ and the bare frequency of the transition. Here we have assumed that this beam does not introduce any additional ac Stark-shifts. Transforming $\hat{H}_{\text{axial}}$ to the qubit frame of reference under $\hat{H}_{\text{sd}}+\hat{H}_{\text{ss}}$, gives the interaction Hamiltonian for the axial beam,

\begin{equation}
   \hat{H}^n_{\text{axial}} = \frac{\Omega}{2} (\sigma_+ e^{-i \delta_n t} + \text{h.c.}), 
\end{equation}
with $\delta_n = \delta - ((1/2)+n)\omega_t-\Delta_0$ the phonon-dependent detuning for an ion located in the node, and $\delta_n = \delta - ((1/2)+n)\omega_t-\Delta_0 -\Delta'_{z} $ for an ion located in the anti-node. When simultaneously addressing the ion with the optical lattice and the axial beam, the ion will experience Rabi oscillations given by,

\begin{equation}
  P_\downarrow(n) =\frac{\Omega}{\Omega_{\text{eff}}} \cos^2(\Omega_{\text{eff}}t / 2),  
\end{equation}
with 
$\Omega_{\text{eff}}(n)^2 = \Omega^2 +\delta_n^2$
the effective Rabi frequency. If we apply the pulse for a $\pi$-time
$t_\pi = \pi/\Omega$
we obtain,

\begin{equation}
    P_\downarrow(n, \delta) =\frac{\Omega}{\Omega_{\text{eff}}} \cos^2(\Omega_{\text{eff}}t_\pi / 2).
\label{eq:n-spectroscopy}
\end{equation}

\begin{figure}
    \centering   \includegraphics[width=\linewidth]{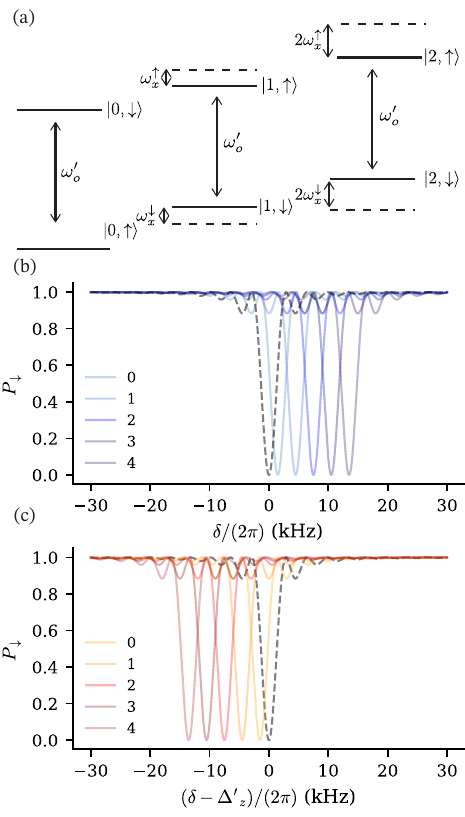}
    \caption{Fock-state-dependent frequency-shift of the qubit. (a) Pictorial representation of the effect of $\hat{H}_{\text{ss}}$ on the bare energy levels of the system. (b) Population in the $\ket{\downarrow}$ state as a function of the detuning $\delta$ from the bare qubit frequency. Different lines show the frequency shift for different Fock numbers when the lattice is on. Dashed lines show the spectroscopy when the lattice is off. (c) Similar to (b), but for an ion sitting in the anti-node of the Standing wave.}
    \label{fig:supplementary-fock-states}
\end{figure}

\aref{fig:supplementary-fock-states} shows the Fock state dependent qubit frequency. \aref{fig:supplementary-fock-states} (a) is a is a pictorial illustration of the state-dependent effect. Here, $\omega'_0 = (1/2)\omega_t+\Delta_0$ is the qubit frequency when the ion is in the motional ground state. Under $\hat{H}_{\text{ss}}$, the $\ket{\downarrow}$ state is downshifted by $n\omega^{\downarrow}_x$ and the $\ket{\uparrow}$ state is upshifted by $n\omega^{\uparrow}_x$.
\aref{fig:supplementary-fock-states} (b) and (c) show \aref{eq:n-spectroscopy}, for the node and the anti-node.
Here we have used $t_\pi = \SI{300}{\micro\second}$, $\omega_t = \SI{3}{\kilo\hertz}$ and $\Delta_0 = 0$. When an ion is located in the node, higher Fock states result in the ion sampling higher intensities for the lattice, and therefore having a higher effective ac Stark-shift. The opposite happens for an ion in the anti-node.

When the ion is not in a Fock state, \aref{eq:n-spectroscopy} is averaged over the phonon occupation distribution, leading to

\begin{equation}
    \label{eq:thermal_carrier_sup}
    P_\downarrow = \sum_{n=0}^{\infty} P_{\bar n}(n) \frac{\Omega}{\Omega_{\text{eff}}} \cos^2(\Omega_{\text{eff}}t_\pi / 2).
\end{equation}

In the case of a thermal state the phonon occupation is

\begin{equation}
    P_{\bar n}(n) = \frac{\bar{n}^n}{(\bar{n} + 1)^{(n + 1)}},
\end{equation}
with mean phonon occupation $\bar{n}$. For a displaced thermal state the distribution is given by

\begin{equation}
\label{eq:populations_thermal_coherent}
        P_{\bar{n},|\alpha|^2}(n) = \sum_{k=0}^\infty \frac{\bar{n}^k}{(\bar n + 1)^{k + 1}} |\braket{n | \alpha, k}|^2, 
\end{equation}
with

\begin{equation}
        |\braket{n | \alpha, k}|^2 = e^{-|\alpha|^2}|\alpha|^{2(n + k)}n!k!\, \left| \sum_{l=0}^k \frac{(-1)^l|\alpha|^{-2l}}{l!(n-l)!(k-l)!} \right|^2,
\end{equation}
with $|\alpha|$ the amplitude of the coherent state \cite{ruster_experimental_2014}. For \aref{fig:distributions} in the main text, we fit \aref{eq:thermal_carrier_sup} using both distributions and with $\bar{n}$, $|\alpha|^2$, and $\Delta_0$ as free parameters. 

\section{Experimental apparatus and sequences}

\subsection{Experimental apparatus}
\begin{figure*}
    \centering
    \includegraphics[width =\textwidth]{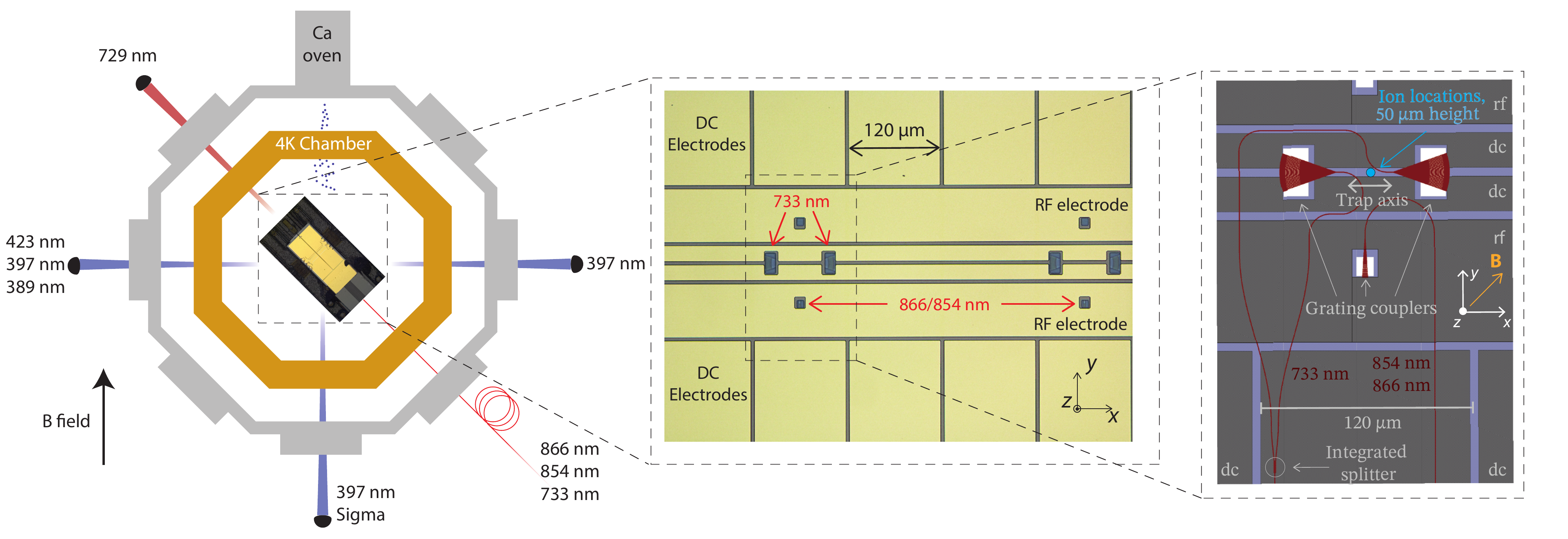}
    \caption{Experimental setup. The trap is located in a 6 K cryostat. Blue light at \SI{389}{\nano\meter}, \SI{397}{\nano\meter} and \SI{423}{\nano\meter} is sent via free-space optics to the loading zone of the trap. Two additional \SI{397}{\nano\meter} beams are sent to the working zone for cooling state preparation and detection. Repumping light at $\SI{854}{\nano\meter}$ and $\SI{866}{\nano\meter}$ is delivered to both zones using integrated waveguides. The lattice at \SI{733}{\nano\meter} is generated from a single waveguide input which is split near the ion using an integrated splitter. After splitting, the light is emitted from two outcouplers located along the trap axis.}
    \label{fig:experimental-setup}
\end{figure*}

Our experimental setup consists in a 6K cryostat, hosting a surface-electrode trap with integrated waveguides and outcouplers (\aref{fig:experimental-setup}). Our trap contains three identical zones with integrated light delivery, out of which we use one for these experiments, and another one for ion loading. Both zones have integrated \SI{866}{\nano\meter} and \SI{854}{\nano\meter} as well as a free-space \SI{397}{\nano\meter} beam. Additionally we deliver free-space light at \SI{423}{\nano\meter} and \SI{389}{\nano\meter} to the loading zone for photoionisation of neutral atoms. Ions from the loading zone are then transported to the working zone. For the case of a two ion crystal, we merge the two ions into a single crystal in the region between the two zones. Having a dedicated loading zone, separated from the working zone, is essential to mitigate stray fields induced from the photoionization lasers and therefore keeping the ions well localised in the optical lattice \cite{ricci_standing-wave_2023}. 

On the working zone we also have integrated light at \SI{733}{\nano\meter}, which is generated using a Titanium-Sapphire laser. This light is fed into the trap from a single input, and we use an integrated waveguide splitter near the trapping location to generate the two beams forming the lattice. Additionally, we use a \SI{397}{\nano\meter} $\sigma-$polarised beam for state preparation and EIT cooling. Finally, we use an ultra-narrow-linewidth free-space axial beam at \SI{729}{\nano\meter} to drive the qubit transition.

\subsection{Experimental sequences}

\begin{figure*}
    \centering
    \includegraphics[width =\textwidth]{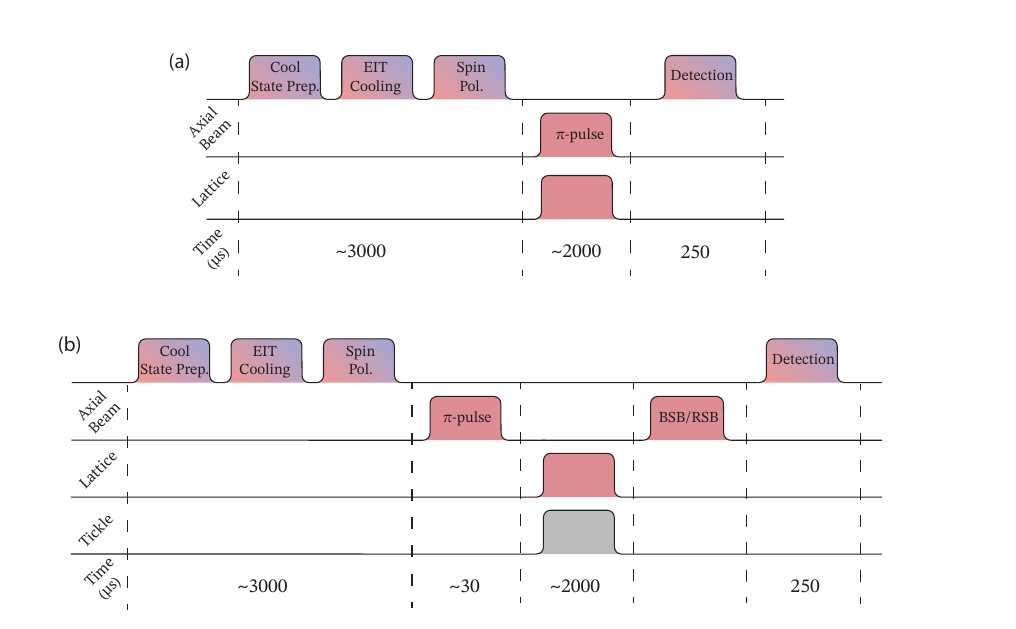}
    \caption{Experimental sequences employed during the experiments. In both sequences the ion is first doppler cooled, followed by EIT cooling and finally optically pumped to the $\ket{\downarrow}$ state using a $\sigma$-polarised 397 beam. At the end of the sequence the ion state is detected using \SI{397}{\nano\meter} fluorescence. (a) Pulse sequence used to perform qubit spectroscopy when the ion is addressed with the lattice. (b) Pulse sequence applied for measuring the state-dependent trapping frequency. A tickle pulse is applied simultaneously to the optical lattice, followed by a red sideband (RSB) pulse for an ion prepared in the $\ket{\downarrow}$ state. Alternatively the ion is prepared in the $\ket{\uparrow}$ state using a $\pi$-pulse with the axial beam. After the tickle pulse a (BSB) probe pulse is applied. }
    \label{fig:pulse-sequences}
\end{figure*}

Our experimental sequences begin by Doppler cooling using the \SI{397}{\nano\meter} as well as the integrated \SI{866}{\nano\meter} and \SI{854}{\nano\meter}. This is followed by an EIT cooling step using both \SI{397}{\nano\meter} beams in the experimental zone and then by a \SI{397}{\nano\meter} $\sigma$ pulse for optically pumping into the $\ket{\downarrow}$ state. We discriminate between the $\ket{\downarrow}$ and $\ket{\uparrow}$ states by collecting \SI{397}{\nano\meter} fluorescence with a PMT for $\SI{250}{\micro\second}$. \aref{fig:pulse-sequences} shows the pulse sequences used in our experiment. \aref{fig:pulse-sequences} (a) shows the pulse sequence used for the experiments ilustrated in \aref{fig:first} (c), \ref{fig:sinlge-ion} (a) and \ref{fig:distributions} (b) of the main text. \aref{fig:pulse-sequences} (b) shows the pulse sequence used for the results shows in \aref{fig:sinlge-ion} (b-d) and \aref{fig:two-ions}. 

\section{Two ion operation}

\subsection{Spacing and trap frequencies}

Working with two ions requires controlling  the position of the center-of-mass of the crystal as well as the spacing between the two ions. In order to maximise the trap frequency shifts, we work in a configuration where both ions are simultaneously in the node or anti-node of the the optical lattice. For a two ion crystal, the separation between the ions is given by 

\begin{equation}
    d(\omega_x) = \left(\frac{e^2}{2\pi\epsilon_0m\omega_x^2}\right)^{1/3},
\end{equation}
with $e$ the fundamental charge \cite{james_quantum_1998}. Matching this spacing to an integer number $j$ of the wavelength of the lattice gives,

\begin{equation}
    d= \frac{j\lambda}{\sin{\theta}}. 
\end{equation}

\begin{figure}
    \centering   \includegraphics[width=\linewidth]{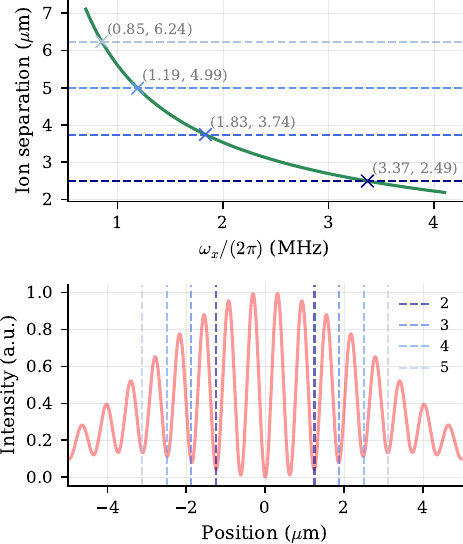}
    \caption{Two-ion positioning in the optical lattice. (a) Ion separation as a function of the trap frequency. Horizontal lines are the distances for which the ion spacing matches the lattice periodicity and the numbers of in parenthesis are the frequencies and separations which match this condition. (b) Equilibrium position of the two ions in the optical lattice for different values of $j$.}
    \label{fig:supplementary-spacing}
\end{figure}

\aref{fig:supplementary-spacing} (a) shows the the ion separation as a function of the trap frequency. Horizontal lines show the ion separation for 4 different values of $j$. The intersection between these lines and $d(\omega_x)$ gives the possible working trap frequencies. \aref{fig:supplementary-spacing} (b) shows a sketch of the intensity landscape of the lattice, created by using $k_x$ from the main text and $x_0=0$. The vertical dashed lines show the equilibrium position of the ions for the different values of $j$. In our experiments we chose $j=3$, where both ions sample a sufficiently large intensity. 

\begin{figure}
    \centering   \includegraphics[width=\linewidth]{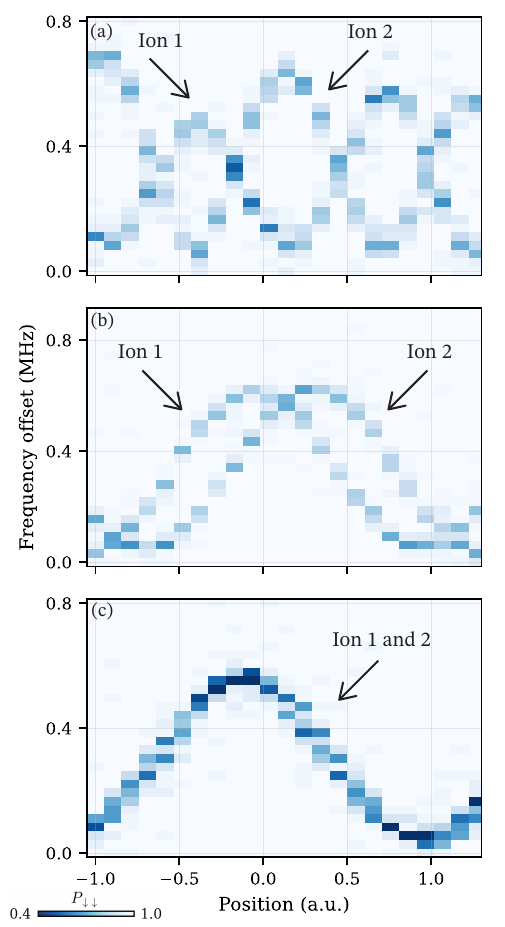}
    \caption{Intensity balancing of two ions. All subfigures show the population in the $\ket{\downarrow\downarrow}$ state as a function of axial position, and the frequency offset of the axial beam. (a) shows the signal when the two ions are sampling different intensities. This pattern is close to one ion being in the anti-node of the lattice while the other is in the node. (b) Intermediate state where both ions don't sit simultaneously in the node or anti-node. (c) Final state where both ions sit simultaneously in the node or the anti-node.}
    \label{fig:supplementary-balancing}
\end{figure}

\begin{figure}
    \centering   \includegraphics[width=\linewidth]{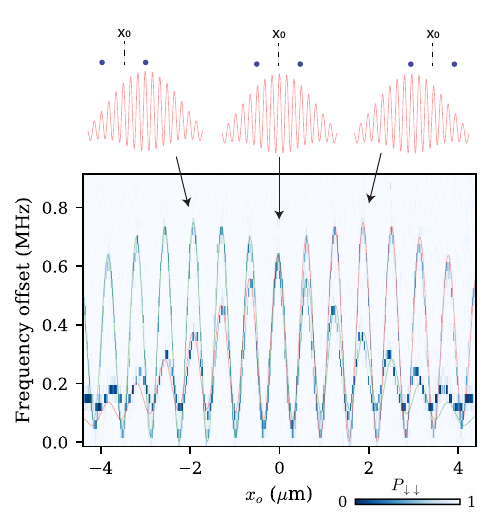}
    \caption{Two ion ac Stark-shift as a function of axial position. The green and red lines are guides to the eye showing the ac Stark-shift sampled by each of the ions. The cartoons on top illustrate the relative position of the ions with respect to the lattice for different equilibrium positions. }
    \label{fig:two-ion-ss}
\end{figure}

\subsection{Trap frequency fine-tuning}

We simultaneously measure the ac Stark-shift in both ions to determine their spacing. For this, we measure the population in the $\ket{\downarrow\downarrow}$ as a function of the equilibrium position of the crystal and the frequency detuning of the \SI{729}{\nano\meter} beam. This population will decrease if either of the two ions absorbs \SI{729}{\nano\meter} light. Therefore, for each equilibrium position of the crystal, we can observe the ac Stark-shift and consequently the intensity on both ions.
\aref{fig:supplementary-balancing} shows the two-ion ac Stark-shift as a function of the equilibrium position, for three different separations between the ions. \aref{fig:supplementary-balancing} (a) shows the case where one ion is close to the node while the second one is close to the anti-node. As we move the equilibrium position, the ion located in the node moves to the anti-node and vice-versa. \aref{fig:supplementary-balancing} (b) shows a measurement where both ions are sampling similar intensities, but are not exactly balanced. \aref{fig:supplementary-balancing} (c) shows the case where both ions sample the same intensity, which is the configuration chosen for the measurements in the main text. 

Once we have found the correct spacing between the ions, we proceed to measure the ac Stark shift for both ions as a function of the axial position in a range of $\pm \SI{4}{\micro\meter}$. The result for this measurement is shown in \aref{fig:two-ion-ss}. The figure shows the population in the $\ket{\downarrow\downarrow}$ state as a function of of the equilibrium position of the two-ion crystal $x_0$ and the detuning of the axial beam from the bare qubit frequency. We can observe that both ions stay simultaneously either in the node or the anti-node independently of the $x_0$. 

\section{Maximal attainable phonon-dependent shifts}

In this section we study the prospects of maximising the state-dependent motional frequency with realistic experimental parameters.
Maximising $\omega_t \propto  \gamma\eta^2I_0$ can be done by maximizing either the available intensity at the ion, the differential polarisability between the qubit states or the Lamb-Dicke parameter. Increasing any of these values can result in negative side effects that need to be taken into account. 

The available intensity at ion location can be increased by reducing the waveguide, incoupling, and grating losses. For example, the grating losses can be reduced by having gratings with increased directivity towards the ion as opposed to the substrate \cite{beck_grating_2024}. Nevertheless, realistically the total losses can hardly be improved by more than a factor of $\approx 2$. The total power into the chip could be increased further, with increased risk of damaging the chip and increased thermal load. The intensity at the ion locations can also be increased by having beams with a tighter focus, however, when trying to address two ions, this can result in a reduced intensity for each of the ions. Alternatively, the use of non-Gaussian beams, where each of the ions is located in an intensity maximum can alleviate these restrictions \cite{beck_grating_2024}. 

\begin{figure}
    \centering
    \includegraphics[width =\linewidth]{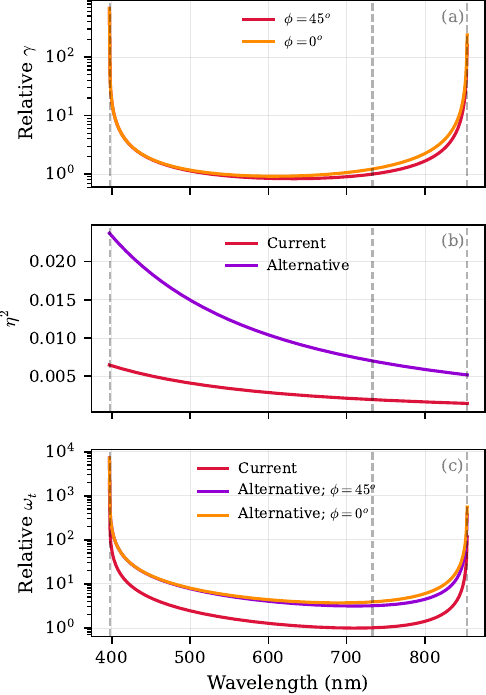}
    \caption{Attainable values of $\gamma$ and $\omega_t$ compared to those in these manuscript for different lattice wavelengths. (a) Relative $\gamma$, where the red line is for the current angle between the polarisation of the lattice and the magnetic field, and yellow is for a configuration where these are parallel. (b) Squared Lamb-Dicke parameter, where the red value is using the trap frequency and emisisson angle from the main text and purple is for a trap frequency of $(2\pi)\times\SI{1}{\mega\hertz}$ and an emission angle of \SI{60}{\degree}. (c) Relative values of $\omega_t$, computed from subfigures (a) and (b). For all plots the dashed lines are at \SI{397}{\nano\meter}, \SI{733}{\nano\meter} and \SI{854}{\nano\meter}. }
    \label{fig:alternative-values}
\end{figure}

A more suitable choice of wavelength for the lattice, as well as an optimised configuration of the experimental parameters can drastically increase the attainable value for $\omega_t$.
Choosing a wavelength near the dipole allowed transitions at \SI{397}{\nano\meter} or \SI{854}{\nano\meter} maximises the $\gamma$ by maximising $\alpha_\downarrow$ or $\alpha_\uparrow$, respectively. In the first case there is also an additional increase from having an increased Lamb-Dicke parameter, while the opposite happens for a lattice at \SI{854}{\nano\meter}.
On the other hand, a lattice at \SI{854}{\nano\meter} can be more easily integrated with silicon nitride  waveguides and losses are expected to be lowered. 
Reducing the trap frequency and designing gratings where the emission angle is lower can also significantly impact the attainable values of $\omega_t$, by increasing the Lamb-Dicke factor. This however would require a reduction in the trap heating rates as well as careful design of the chip layout.

\aref{fig:alternative-values} shows the relative value for $\gamma$ and $\omega_t$ for different lattice wavelengths as well as different configurations.
\aref{fig:alternative-values} (a) shows the relative value of $\gamma$ to the current value as a function of wavelength.
The red curve is for the same magnetic field configuration as the current one, whereas the orange curve is for a configuration where the magnetic field is parallel to the polarisation of the light, where an increase in the polarisability of the $\ket{\uparrow}$ state can be achieved. 
\aref{fig:alternative-values} (b) shows $\eta^2$ for the when setting $\omega_x = (2\pi)\times\SI{1.46}{\mega\hertz}$ and an emission angle of $\theta = \SI{36}{\degree}$ in red and $\omega^{(a)}_x = (2\pi)\times\SI{1}{\mega\hertz}$ and an emission angle of $\theta^{(a)} = \SI{60}{\degree}$ in purple.
\aref{fig:alternative-values} (c) shows the value of $\omega_t$ for different wavelength and configurations, relative to the current value. Purple and orange curves are created using the alternative parameters $\omega^{(a)}_x$ and $\theta^{(a)}$. 
As an example, for a lattice at \SI{850}{\nano\meter}, and using the alternative parameters and $\phi=0$ we expect an increase in $\gamma$ by a factor of $\sim 24$, and an increase in $\omega_t$ of $\sim 56$. For a lattice at \SI{400}{\nano\meter} we expect an increase in $\gamma$ of $\sim 17$ and an increase in $\omega_t$ of $\sim 185$. These values can be increased even further by having the lattice wavelength closer to the dipole transitions at $\SI{854}{\nano\meter}$ and $\SI{397}{\nano\meter}$, at the expense of increased scattering rates. 

The use of a lattice to create the optical potential as opposed to an optical tweezer has the advantage of reducing the scattering rates for an ion located in the node. This would allow in principle to operate the lattice near the dipole-allowed transitions while keeping low scattering rates. The scattering rates would be fundamentally limited by the intensity sampled by the ion in the motional ground state. In this sense, the lattice allows for a suppression of the scattering rates in the order of $\eta^2$ (\aref{fig:alternative-values} (b)), when locating the ion in the node as opposed to the anti-node. For the alternative parameters $\omega^{(a)}_x$ and $\theta^{(a)}$ this suppression corresponds to a factor of $\sim 43$ and $\sim 193$ for lattices at \SI{400}{\nano\meter} and \SI{850}{\nano\meter}, respectively. In practice, the scattering rates could be higher due to imperfect power cancellation in the nodes.

\bibliography{bibliography}

\end{document}